\documentstyle[aps,eqsecnum]{revtex}
\begin{document}

\draft
\title{Kaluza-Klein Formalism of General Spacetimes}
\author{ J.H. Yoon\thanks{Electronic address: 
yoonjh@cosmic.konkuk.ac.kr}}
\address{Department of Physics\\
Konkuk University, Seoul 143-701, Korea}

\maketitle

\begin{abstract}
I describe the Kaluza-Klein approach to general relativity 
of 4-dimensional spacetimes. This approach is based 
on the (2,2)-fibration of a generic 4-dimensional spacetime, 
which is viewed as a local product of 
a (1+1)-dimensional base manifold and a 2-dimensional fibre 
space. It is shown that the metric coefficients can be decomposed
into sets of fields, which transform as a tensor field,
gauge fields, and scalar fields with respect to 
the infinite dimensional group of the diffeomorphisms 
of the 2-dimensional fibre space. I discuss a few applications
of this formalism.

\end{abstract}


\begin{section}{Introduction}
\label{s1}
It has been known for some time that there is a curious 
correspondence between (self-dual) Yang-Mills equations and
the (self-dual) Einstein's equations, when the Yang-Mills 
gauge symmetry is extended to an infinite dimensional symmetry 
of (volume-preserving) diffeomorphisms of some auxiliary 
manifold\cite{mason-newman89}. It is also well-known that
the equations of motion of 2-dimensional non-linear 
sigma models with the target space as the area-preserving 
diffeomorphism of an auxiliary 
2-surface\cite{park90,park92,bakas89,floratos89,fairlie90} 
are identical to the the self-dual Einstein's equations
written in the Pleba\~{n}ski form\cite{plebanski75}. 

These correspondences 
are most striking for self-dual cases, and indicate
an intriguing possibility that we may be 
able to reconstruct the full Einstein's general relativity from suitable 
gauge field theories by replacing the usual finite dimensional 
gauge symmetry with an infinite dimensional group of the diffeomorphisms
of some manifold. If we recall that the gauge symmetry of 
general relativity is the group of the diffeomorphisms 
of a 4-dimensional spacetime, this seemingly wild 
speculation is not totally unreasonable. Recently we have shown that 
such a description is indeed possible, by rewriting
the Einstein-Hilbert action of general relativity of 
generic 4-dimensional spacetimes 
in the (2,2)-decomposition
\cite{yoon92,yoon93a,yoon93b,dinverno78,dinverno80,hayward93}. 
In this approach, the 4-dimensional 
spacetime is viewed, at least for a finite range of the spacetime,
as a locally fibred manifold that consists of a 
(1+1)-dimensional base manifold $M_{1+1}$ and a 2-dimensional 
fibre space $N_{2}$.

The Yang-Mills gauge fields, which naturally appear in this 
Kaluza-Klein setting\cite{cho75}, 
are defined on the (1+1)-dimensional base manifold $M_{1+1}$, 
and turn out to be valued 
in the Lie algebra of an infinite dimensional group
of the diffeomorphisms of the 2-dimensional 
fibre space $N_{2}$ (i.e. diff$N_{2}$).
This feature is expected to simplify considerably 
certain issues concerned with the constraints of 
general relativity. Namely,
in Yang-Mills gauge theories, it is well-known that 
the Gauss-law constraints associated with the Yang-Mills 
gauge invariance can be made ``trivial'', 
if we consider gauge invariant quantities only.
Thus, in principle, one might expect that
the problem of solving  the constraints of general relativity 
could be made ``trivial'', 
at least for some of them, if such a gauge theory description 
is possible. 
The purpose of this paper is to show explicitly that 
our variables transform as a tensor field, 
gauge fields, and scalar fields with respect to 
the diff$N_{2}$ transformations, 
and discuss a general spacetime
from the 4-dimensional fibre bundle point of view.

This paper is organized as follows. 
In section \ref{s2}, we shall outline 
the kinematics of the (2,2)-decomposition 
of a generic 4-dimensional spacetime, and introduce 
the Kaluza-Klein (KK) variables
{\it without} assuming any spacetime isometries.
In section \ref{s3}, we shall find the transformation properties 
of the KK variables with respect to the diff$N_{2}$ transformations,
and introduce the notion of the diff$N_{2}$-{\it covariant} derivatives.
In section \ref{s4}, we shall write down 
the Einstein-Hilbert action, and finally, 
we discuss possible applications of this formalism.

\end{section}

\begin{section}{Kinematics}
\label{s2}

Let us decompose a generic 4-dimensional spacetime
of the Lorentzian signature from the KK perspective, 
in which the spacetime under consideration 
is viewed as a 4-dimensional fibre bundle, consisting 
of a (1+1)-dimensional base manifold $M_{1+1}$ and 
a 2-dimensional fibre space $N_{2}$.
Let the basis vector fields of  $M_{1+1}$ and $N_{2}$ be
$\partial / \!  \partial x^{\mu}(=\partial_{\mu})$
and $\partial / \!  \partial y^{a}(=\partial_{a})$,
respectively, where $\mu=0,1$ and $a=2,3$.
The horizontal vector fields $\hat{\partial}_{\mu}$, 
which are defined to be orthogonal to $N_{2}$, 
can be expressed as linear combinations
of $\partial_{\mu}$ and $\partial_{a}$, 
\begin{equation}
\hat{\partial}_{\mu}=\partial_{\mu} 
- A_{\mu}^{\ a}\partial_{a},                \label{combi}
\end{equation}
where the fields $A_{\mu}^{\ a}$ are functions of
$(x^{\mu},y^{a})$. 
Let us denote by  $\gamma^{\mu\nu}$ the inverse metric 
of the horizontal space spanned by $\hat{\partial}_{\mu}$, 
and by $\phi^{a b}$ the inverse metric of $N_{2}$,  respectively.
In the horizontal lift basis which consists of 
$\{ \hat{\partial}_{\mu}, \partial_{a} \}$,
the metric of the 4-dimensional spacetime 
can then be written as\cite{mtw72}
\begin{equation}
\Big( {\partial \over \partial s}\Big)^{2}=\gamma^{\mu\nu}
\Big( \partial_{\mu} - A_{\mu}^{\ a}\partial_{a} \Big)
\otimes
\Big( \partial_{\nu} - A_{\nu}^{\ b}\partial_{b} \Big)
+\phi^{a b}\partial_{a}\otimes \partial_{b}.  \label{dualmetric}
\end{equation}
In the corresponding dual basis
$\{ dx^{\mu}, dy^{a} + A_{\mu}^{\ a}dx^{\mu} \}$,
the metric becomes 
\begin{equation}
ds^{2}=\gamma_{\mu\nu}dx^{\mu}dx^{\nu} + \phi_{a b}
\Big( dy^{a} + A_{\mu}^{\ a} dx^{\mu} \Big)
\Big( dy^{b} + A_{\nu}^{\ b} dx^{\nu} \Big).  \label{metric}
\end{equation}
Formally the above metric looks similar to the
``dimensionally reduced" metric in standard KK theories,
but in fact it is quite different.
In the standard KK reduction certain isometries are usually assumed, 
and dimensional reduction is made by projection
along the directions generated by these isometries\cite{cho75}. 
There the fields $A_{\mu}^{\ a}$ are identified 
as the KK gauge fields associated with the {\it finite} dimensional 
isometry group. In this paper, 
we do {\it not} assume such isometries: nevertheless, 
it turns out that the KK idea still works, 
and as we shall show shortly, the fields  
$A_{\mu}^{\ a}$ can be identified as the gauge fields
valued in the {\it infinite} dimensional Lie algebra of 
the diff$N_{2}$ transformations.
Moreover, the fields $\phi_{a b}$ and $\gamma_{\mu\nu}$ 
transform as a tensor field and scalar fields with respect to 
the diff$N_{2}$ transformations.
\end{section}

\begin{section}{Diffeomorphisms as a local gauge symmetry}
\label{s3}
\begin{subsection}{Finite transformations}

Let us find the transformation properties of the fields 
$\phi_{a b}$, $A_{\mu}^{\ a}$,  and $\gamma_{\mu\nu}$ 
with respect to the diff$N_{2}$ transformations,
which are the following coordinate transformations of $N_{2}$,
while keeping $x^{\mu}$ constant\cite{exact80},
\begin{equation}
y^{' a}=y^{' a} (x,y), 
\hspace{.5cm}
x^{' \mu}=x^{\mu}.                      \label{new}
\end{equation}
Thus we have
\begin{equation}
dy^{a}={\partial y^{a}\over \partial y^{'c}}
\Big\{ dy^{'c} - \Big( {\partial y^{'c}\over \partial x^{\mu}}\Big)
  dx^{'\mu} \Big\},   
\hspace{.5cm}
dx^{\mu}=dx^{' \mu}.            \label{newb}
\end{equation}
In the new coordinates the term proportional to
$dx^{\mu}dy^{a}$ in (\ref{metric}) becomes, 
keeping the $(x^{\mu},y^{a})$ dependence explicit,
\begin{eqnarray}
& &2\phi_{a b}(x,y)A_{\mu}^{\ a}(x,y)dx^{\mu}dy^{b} \nonumber\\
&=&2\Big( {\partial y^{a}\over \partial y^{'c}} \Big)
 \Big( {\partial y^{b}\over \partial y^{'d}} \Big) \phi_{a b}(x,y)
\Big(  {\partial y^{'d}\over \partial y^{e}} \Big)
A_{\mu}^{\ e}(x,y)dx^{'\mu}
\Big\{ dy^{'c} -
\Big( {\partial y^{'c} \over \partial x^{\nu}}\Big)
dx^{'\nu} \Big\},                   \label{newd}
\end{eqnarray}
where the identity
\begin{equation}
\Big( {\partial y^{a}\over \partial y^{'d}} \Big)
\Big( {\partial y^{'d}\over \partial y^{e}} \Big)
=\delta_{e}^{\ a}
\end{equation}
was used. Also
the term proportional to $dy^{a}dy^{b}$ becomes
\begin{eqnarray}
& &\phi_{a b}(x,y)dy^{a}dy^{b}            \nonumber\\
&=&\Big(   {\partial y^{a}\over \partial y^{'c}}\Big)
\Big( {\partial y^{b}\over \partial y^{'d}}\Big) \phi_{a b}(x,y)
  \Big\{
  dy^{'c} dy^{'d}
  -2\Big( {\partial y^{'d} \over \partial x^{\mu}}\Big) 
  dy^{'c}dx^{' \mu}
 +\Big( {\partial y^{'c}\over \partial x^{\mu}}\Big)
 \Big( {\partial y^{'d}\over \partial x^{\nu}}\Big)
  dx^{'\mu}dx^{'\nu}\Big\}.            \label{newc}
\end{eqnarray}
After rearranging terms, the metric (\ref{metric}) 
can be written as, in the new coordinates,
\begin{eqnarray}
ds^{2}
&=&\gamma_{\mu\nu}(x,y)dx^{'\mu}dx^{'\nu}
  +\Big( {\partial y^{a}\over \partial y^{'c}}\Big)
  \Big( {\partial y^{b}\over \partial y^{'d}}\Big)
  \phi_{a b}(x,y) dy^{'c} dy^{'d} \nonumber\\
& &+2\Big( {\partial y^{a}\over \partial y^{'c}}\Big)
  \Big( {\partial y^{b}\over \partial y^{'d}}\Big)
  \phi_{a b}(x,y)
  \Big\{ 
   \Big( {\partial y^{' d}\over \partial y^{e}}\Big) 
  A_{\mu}^{\ e}(x,y)
  -{\partial y^{' d}\over \partial x^{\mu}} \Big\}
   dx^{'\mu}dy^{'c}\nonumber\\
& &+\phi_{a b}(x,y)  \Big\{
 A_{\mu}^{\ a}(x,y)A_{\nu}^{\ b}(x,y)
- 2\Big( {\partial y^{a}\over \partial y^{'c}}\Big)
  \Big( {\partial y^{b}\over \partial y^{'d}}\Big)
  \Big( {\partial y^{' d}\over \partial y^{e}} \Big) 
  A_{\mu}^{\ e}(x,y)
   \Big( {\partial y^{' c}\over \partial x^{\nu}}  \Big) \nonumber\\
& &+ \Big( {\partial y^{a}\over \partial y^{'c}}\Big)
  \Big( {\partial y^{b}\over \partial y^{'d}}\Big)
 \Big( {\partial y^{' c}\over \partial x^{\mu}} \Big)
 \Big( {\partial y^{' d}\over \partial x^{\nu}} \Big)
\Big\} dx^{'\mu}dx^{'\nu},              \label{large}
\end{eqnarray}
which must be equal to
\begin{equation}
ds^{'2}=\gamma'_{\mu\nu}(x', y')dx^{'\mu}dx^{'\nu}
+ \phi'_{a b}(x', y')\Big\{ dy^{'a} + A_{\mu}^{'\ a}(x', y')
dx^{'\mu} \Big\}
\Big\{ dy^{'b} + A_{\nu}^{'\ b}(x', y')
dx^{'\nu} \Big\},                           \label{newline}
\end{equation}
since the line element is invariant under the diff$N_{2}$ 
transformations.
If we compare terms containing $dy^{'a}dy^{'b}$, we find that
$\phi_{a b}(x, y)$ transform as
\begin{equation}
\phi'_{a b}(x', y')
=\Big( {\partial y^{c}\over \partial y^{'a}}\Big)
 \Big( {\partial y^{d}\over \partial y^{'b}}\Big)
  \phi_{c d}(x,y).                         \label{coeff1}
\end{equation}
This shows that $\phi_{a b}(x,y)$ is a tensor field
with respect to the diff$N_{2}$ transformations.
If we use the equation (\ref{coeff1}) in (\ref{large}), 
the metric becomes 
\begin {eqnarray}
ds^{2}&=&\gamma_{\mu\nu}(x, y)dx^{'\mu}dx^{'\nu}
 +\phi'_{c d}(x', y')dy^{' c}dy^{' d} 
+2\phi'_{c d}(x', y')\Big\{
 \Big( {\partial y^{' d}\over \partial y^{a}}\Big) A_{\mu}^{\ a}(x, y)
  -{\partial y^{' d}\over \partial x^{\mu}}
   \Big\}
   dx^{'\mu}dy^{'c}               \nonumber\\
& &+\phi'_{c d}(x', y')
   \Big\{
 \Big( {\partial y^{' c}\over \partial y^{a}}\Big) A_{\mu}^{\ a}(x, y)
  -{\partial y^{' c}\over \partial x^{\mu}} \Big\}
  \Big\{
 \Big( {\partial y^{' d}\over \partial y^{b}} \Big) A_{\nu}^{\ b}(x, y)
  -{\partial y^{' d}\over \partial x^{\nu}}\Big\} 
  dx^{'\mu}dx^{'\nu},
\end{eqnarray}
from which we deduce the following transformation properties of 
$A_{\mu}^{\ a}(x, y)$ and $\gamma_{\mu\nu}(x,y)$
\begin{eqnarray}
& &A_{\mu}^{'\ a}(x', y')
=\Big( {\partial y^{'a}\over \partial y^{b}}\Big) A_{\mu}^{\ b}(x, y)
-{\partial y^{'a}\over \partial x^{\mu}}(x, y), \label{coeff2}\\
& &\gamma'_{\mu\nu}(x',y')=\gamma_{\mu\nu}(x,y), \label{coeff3}
\end{eqnarray}
under the diff$N_{2}$ transformations. 
\end{subsection}

\begin{subsection}{Infinitesimal transformations}

It will be instructive to examine the infinitesimal 
transformations corresponding to 
the above finite diff$N_{2}$ transformations.
The infinitesimal diff$N_{2}$ transformations
consist of the following transformations
\begin{equation}
y^{'a}=y^{a} + \xi^{a}(x,y),
\hspace{.5cm} 
x^{' \mu}= x^{\mu} 
\hspace{.5cm} ({\rm O}(\xi^{2}) \ll 1),  \label{infff}
\end{equation}
where $\xi^{a}(x,y)$ is an arbitrary, infinitesimal, function
of $(x^{\mu}, y^{a})$.
From this it follows that 
\begin{equation}
{\partial y^{c}\over \partial y^{'a}}
=\delta_{a}^{\ c}
- {\partial \xi^{c}\over \partial y^{a}}
 + \cdots,                      \label{idd}
\end{equation}
where $\cdots$ means terms of ${\rm O}(\xi^{2})$.
If we expand the l.h.s. of the equation (\ref{coeff1}) 
in $\xi^{a}$, it becomes
\begin{equation}
\phi'_{a b}(x',y+\xi)=\phi'_{a b}(x,y)
+ \xi^{c}{\partial \over \partial y^{c}}\phi_{a b}(x,y)
+ \cdots,                  \label{left}
\end{equation}
whereas the r.h.s. becomes
\begin{eqnarray}
\Big( {\partial y^{c}\over \partial y^{'a}}\Big)
\Big( {\partial y^{d}\over \partial y^{'b}}\Big)
\phi_{c d}(x,y)
&=&\phi_{a b}(x,y)-
{\partial \xi^{c}\over \partial y^{a}}\phi_{c b}(x,y)
-{\partial \xi^{c}\over \partial y^{b}}\phi_{a c}(x,y)
+ \cdots.                      \label{heya}
\end{eqnarray}
Thus we have 
\begin{eqnarray}
\delta \phi_{a b}(x,y)
&\equiv &\phi'_{a b}(x,y)- \phi_{a b}(x,y) \nonumber\\
&=&-\xi^{c}\partial_{c} \phi_{a b}(x,y)
   -(\partial_{a} \xi^{c}) \phi_{c b}(x,y)
   -(\partial_{b}\xi^{c})\phi_{a c}(x,y)                    \nonumber\\
&=&-[ \xi, \phi ]_{{\rm L} a b},           \label{finn}
\end{eqnarray}
where the subscript ${}_{\rm L}$ denotes 
the Lie derivative along the vector field 
$\xi \equiv \xi^{a}\partial_{a}$, i.e. 
\begin{equation}
[ \xi, \phi ]_{{\rm L} a b}
=\xi^{c}\partial_{c} \phi_{a b}
   +(\partial_{a} \xi^{c})\phi_{c b}
   +(\partial_{b} \xi^{c})\phi_{a c}.
\end{equation}
It is a straightforward exercise to derive the infinitesimal 
transformation properties $A_{\mu}^{\ a}$ and $\gamma_{\mu\nu}$ 
from (\ref{coeff2}) and (\ref{coeff3}). They are found to be
\begin{eqnarray}
\delta A_{\mu}^{\ a}(x,y)
&=&-\partial_{\mu} \xi^{a}
   + [A_{\mu},\ \xi ]_{{\rm L}}^{a}       \nonumber\\
&=&-\partial_{\mu} \xi^{a} 
   + A_{\mu}^{\ c}\partial_{c} \xi^{a}
    -\xi^{c}\partial_{c}A_{\mu}^{\ a},  \label{hi}\\
\delta \gamma_{\mu\nu}(x,y)
&=&-[ \xi, \gamma_{\mu\nu}]_{{\rm L}}   \nonumber\\
&=&-\xi^{a}\partial_{a}\gamma_{\mu\nu},        \label{hia}
\end{eqnarray} 
where $[A_{\mu},\ \xi ]_{{\rm L}}^{a}$ and 
$[ \xi, \gamma_{\mu\nu}]_{{\rm L}}$ 
are the Lie derivatives of $\xi^{a}$ and
$\gamma_{\mu\nu}$ along the vector fields
$A_{\mu}=A_{\mu}^{\ c}\partial_{c}$ and 
$\xi=\xi^{c}\partial_{c}$, respectively.
Notice that the Lie derivative acts on the fibre space 
index ($a$) only.
The equations (\ref{finn}), (\ref{hi}), and (\ref{hia})
clearly show that the metric components 
$\{ \phi_{ab}, A_{\mu}^{\ a}, \gamma_{\mu\nu}\}$
transform as a tensor field, gauge fields, and scalar fields
under the diff$N_{2}$ transformations, respectively.

\end{subsection}

\begin{subsection}{diff$N_{2}$-covariant derivative}

Using the Lie derivative along the diff$N_{2}$-valued gauge fields,
the diff$N_{2}$-{\it covariant} derivative $D_{\mu}$ 
can be {\it naturally} defined as
\begin{equation}
D_{\mu}=\partial_{\mu}-[A_{\mu}, \ \ ]_{{\rm L}}.       \label{cd} 
\end{equation}
With this definition, 
the equation (\ref{hi}) can be written as 
\begin{equation}
\delta A_{\mu}^{\ a}=-D_{\mu}\xi^{a},
\end{equation}
which suggests that 
the diff$N_{2}$-valued field strength $F_{\mu\nu}^{\ \ a}$ 
be defined as
\begin{equation}
[D_{\mu},D_{\nu}]_{{\rm L}} \eta 
= -F_{\mu\nu}^{\ \ a}\partial_{a}\eta 
\end{equation}
for an arbitrary scalar function $\eta$, 
where $F_{\mu\nu}^{\ \ a}$ is given by
\begin{eqnarray}
F_{\mu\nu}^{\ \ a}
&=&\partial_\mu A_{\nu} ^ { \ a}-\partial_\nu A_{\mu} ^ { \ a} 
- [A_{\mu}, A_{\nu}]_{\rm L}^{a}      \nonumber\\
&=&\partial_\mu A_{\nu} ^ { \ a}-\partial_\nu A_{\mu} ^ { \ a}
-A_{\mu}^{\ c}\partial_{c}A_{\nu}^{\ a}
+A_{\nu}^{\ c}\partial_{c}A_{\mu}^{\ a}.          \label{fiea}
\end{eqnarray}
Similarly, the diff$N_{2}$-covariant derivative
of $\phi_{ab}$ is defined as 
\begin{eqnarray}
D_{\mu}\phi_{a b}
&=&\partial_{\mu}\phi_{a b} 
- [A_{\mu}, \phi ]_{{\rm L} a b}       \nonumber\\
&=&\partial_{\mu}\phi_{a b}
-A_{\mu}^{\ c}\partial_{c}\phi_{a b}
-(\partial_{a}A_{\mu}^{\ c})\phi_{b c}
-(\partial_{b}A_{\mu}^{\ c})\phi_{a c}.       \label{cophi}
\end{eqnarray}

It remains to show that
$F_{\mu\nu}^{\ \ a}$ and $D_{\mu}\phi_{a b}$ 
transform {\it covariantly} under the
infinitesimal diff$N_{2}$ transformations (\ref{infff}).
Let us consider $D_{\mu}\phi_{a b}$ first.  
The infinitesimal transformation of $D_{\mu}\phi_{a b}$ becomes
\begin{equation}
\delta ( D_{\mu}\phi_{a b} )
=-\partial_\mu \Big( [ \xi , \phi]_{{\rm L} ab} \Big)
+[ A_{\mu} ,  [\xi, \phi ]_{\rm L}  ]_{{\rm L} ab}
+[D_{\mu}\xi, \phi ]_{{\rm L} ab},            \label{varcophi}
\end{equation}
where we used the equations (\ref{finn}) and (\ref{hi}),
and the Lie brackets are
\begin{eqnarray}
& &[ A_{\mu} ,  [\xi, \phi ]_{\rm L}   ]_{{\rm L} ab}
=A_{\mu} ^ { \ c}\partial_c \Big( [ \xi , \phi ]_{{\rm L} ab} \Big)
+(\partial_a A_{\mu} ^ { \ c})[ \xi , \phi ]_{{\rm L} bc}
+(\partial_b A_{\mu} ^ { \ c})[ \xi , \phi ]_{{\rm L} ac}, \\
& &[D_{\mu}\xi, \phi ]_{{\rm L} ab}
=(D_{\mu}\xi^{c})( \partial_c \phi_{a b} )
+\partial_a ( D_{\mu}\xi^{c} ) \phi_{ bc} 
+\partial_b ( D_{\mu}\xi^{c} ) \phi_{ ac}.
\end{eqnarray}
Using the Leibniz rule of the derivative $\partial_{\mu}$
\begin{equation}
\partial_\mu \Big( [ \xi , \phi]_{{\rm L} ab} \Big)
=[\partial_\mu  \xi , \phi]_{{\rm L} ab} 
+[ \xi , \partial_\mu \phi]_{{\rm L} ab},
\end{equation}
and the properties of the Lie bracket
\begin{eqnarray}
& &[ D_{\mu}\xi, \phi  ]_{{\rm L} ab}
= [\partial_\mu  \xi ,\phi ]_{{\rm L} ab} 
- [  [A_{\mu}, \xi ]_{\rm L},\phi  ]_{{\rm L} ab}, \\
& &[ A_{\mu} ,  [\xi, \phi ]_{\rm L}  ]_{{\rm L} ab}
=-[\xi, [ \phi , A_{\mu}]_{\rm L}  ]_{{\rm L} ab}
- [\phi , [A_{\mu}, \xi ]_{\rm L}  ]_{{\rm L} ab},
\end{eqnarray}
we find that the equation (\ref{varcophi}) becomes
\begin{eqnarray}
\delta ( D_{\mu}\phi_{a b} )
&=&- [\xi, \partial_{\mu}\phi    ]_{{\rm L} ab} 
   + [ \xi, [ A_{\mu},   \phi  ]_{\rm L}  ]_{{\rm L} ab}  \nonumber\\
&=&- [  \xi, D_\mu \phi ]_{{\rm L} ab},
\end{eqnarray}
which shows that $ D_{\mu}\phi_{a b}$ transforms {\it covariantly} under
the diff$N_{2}$ transformation.

Similarly, the infinitesimal transformation 
$\delta F_{\mu\nu}^{\ \ a}$  becomes
\begin{equation}
\delta F_{\mu\nu}^{\ \ a}
=\partial_\mu \Big( [ A_{\nu}, \xi ]_{\rm L}^{a}  \Big)
+[   D_{\mu}\xi, A_{\nu}  ]_{\rm L}^{a}
-(\mu \leftrightarrow\nu ).           \label{stop}
\end{equation}
Using the following identities 
\begin{eqnarray}
& &\partial_\mu \Big( [ A_{\nu}, \xi ]_{\rm L}^{a}  \Big)
  =[ \partial_\mu A_{\nu}, \xi  ]_{\rm L}^{a}
  +[A_{\nu}, \partial_\mu \xi ]_{\rm L}^{a},  \\
& &[ D_{\mu}\xi, A_{\nu} ]_{\rm L}^{a}
 =-[ A_{\nu}, D_{\mu}\xi  ]_{\rm L}^{a}   
 =-[ A_{\nu}, \partial_\mu \xi  ]_{\rm L}^{a}
 +[ A_{\nu}, [  A_{\mu},  \xi ]_{\rm L}^{} ]_{\rm L}^{a},
\end{eqnarray}
we find that
\begin{eqnarray}
\delta F_{\mu\nu}^{\ \ a}
&=&[\partial_\mu A_{\nu}-\partial_\nu A_{\mu} , \xi ]_{\rm L}^{a}
+[A_{\nu}, [A_{\mu}, \xi   ]_{\rm L}^{}   ]_{\rm L}^{a} 
-[A_{\mu}, [A_{\nu}, \xi   ]_{\rm L}^{}   ]_{\rm L}^{a} \nonumber\\
&=&-[ \xi , F_{\mu\nu}]_{\rm L}^{a},
\end{eqnarray}
where we used the Jacobi identity
\begin{equation}
[ A_{\nu}, [ A_{\mu},  \xi ]_{\rm L}^{}  ]_{\rm L}^{a}
=-[ A_{\mu}, [\xi, A_{\nu}  ]_{\rm L}^{} ]_{\rm L}^{a}
-[  \xi, [ A_{\nu}, A_{\mu} ]_{\rm L}^{} ]_{\rm L}^{a}.
\end{equation}
Therefore it follows that
\begin{equation}
\delta F_{\mu\nu}^{\ \ a}=-[ \xi , F_{\mu\nu}]_{\rm L}^{a},
\end{equation}
which shows that $F_{\mu\nu}^{\ \ a}$ is indeed 
the diff$N_{2}$-valued field strength.

It must be marked here that, in the
(2,2)-KK formalism, 
the Lie derivative, rather than the covariant derivative,
appears naturally.
The appearance of an infinite dimensional symmetry
such as diff$N_{2}$ is not surprising, 
since in general relativity the underlying gauge symmetry is 
the infinite dimensional group of
the diffeomorphisms of a 4-dimensional spacetime.
The point is that it is the diff$N_{2}$ symmetry, 
the subgroup of the diffeomorphisms of a 4-dimensional spacetime, 
that shows up as a local gauge symmetry of the Yang-Mills type.
This implies that  
the (2,2)-KK formalism can be made a viable method 
of studying general relativity
from the standpoint of the (1+1)-dimensional Yang-Mills gauge theory
with the diff$N_{2}$ symmetry as a local gauge symmetry.

\end{subsection}
\end{section}

\begin{section}{The Action}
\label{s4}
The Einstein-Hilbert action in this KK formalism is given by
\begin{eqnarray}
I&=&\int \! \! d^{2}x d^{2}y \,
\sqrt{-\gamma}\sqrt{\phi} \, \Big[
\gamma^{\mu\nu}\hat{ {\rm R}}_{\mu\nu} + \phi^{a c}{\rm R}_{a c}
 + {1\over 4}\gamma^{\mu\nu}\gamma^{\alpha\beta}
  \phi_{a b}F_{\mu\alpha} ^ { \  \ a}
  F_{\nu\beta}^{\ \ b}          \nonumber\\
& & +{1\over 4}\gamma^{\mu\nu}\phi ^ {a b}\phi ^ {c d}\Big\{
  (D_{\mu}\phi_{a c})(D_{\nu}\phi_{b d})
  -(D_{\mu}\phi_{a b})(D_{\nu}\phi_{c d}) \Big\}  \nonumber\\
& &  +{1\over 4}\phi ^ {a b}\gamma^{\mu\nu}
  \gamma^{\alpha\beta}\Big\{
 (\partial_{a}\gamma_{\mu \alpha})(\partial_{b}\gamma_{\nu\beta})
 -(\partial_{a}\gamma_{\mu \nu})(\partial_{b}
  \gamma_{\alpha\beta}) \Big\} \Big] 
+\int \! \! d^{2}x d^{2}y \, (\partial_{A}S^{A}).  \label{big}
\end{eqnarray}
Let us summarize the notations:\\
1. The curvature tensors 
$\hat{ {\rm R}}_{\mu\nu}$ and $ {\rm R}_{a c}$  are defined as
\begin{eqnarray}
& &\hat{ {\rm R}}_{\mu\nu}=
  \hat{\partial}_{\mu}^{}\hat{ \Gamma}_{\alpha \nu}^{\ \ \alpha}
 -\hat{\partial}_{\alpha}^{}
  \hat{ \Gamma}_{\mu \nu}^{\ \ \alpha}
 +\hat{ \Gamma}_{\mu \beta}^{\ \ \alpha}
  \hat{ \Gamma}_{\alpha \nu}^{\ \ \beta}
 -\hat{ \Gamma}_{\beta \alpha}^{\ \ \beta}
  \hat{ \Gamma}_{\mu \nu}^{\ \ \alpha},       \label{gric}\\
& &{\rm R}_{a c}=\partial_{a}^{}\Gamma_{b c}^{\ \ b}
-\partial_{b}^{}\Gamma_{a c}^{\ \ b}
+\Gamma_{a d}^{\ \ b}
 \Gamma_{b c}^{\ \ d}
-\Gamma_{d b}^{\ \ d}\Gamma_{a c}^{\ \ b}, \label{qten}\\
& &\hat{ \Gamma}_{\mu \nu}^{\ \ \alpha}
={1\over 2}\gamma^{\alpha\beta}\Big(
 \hat{\partial}_{\mu}\gamma_{\nu\beta}
 + \hat{\partial}_{\nu}\gamma_{\mu\beta}
 -\hat{\partial}_{\beta}\gamma_{\mu\nu}  \Big),  \label{ok}\\
& &\Gamma_{a b}^{\ \ c}
  ={1\over 2}\phi^{c d}\Big(
  \partial_{a}\phi_{b d} + \partial_{b}\phi_{a d}
 -\partial_{d}\phi_{a b}\Big).   \label{fibre}
\end{eqnarray}
\noindent
2. The last term in (\ref{big}) is a surface integral, where
$S^{A}=(S^{\mu}, S^{a})$ is given by
\begin{eqnarray}
& & S^{\mu}= \sqrt{-\gamma}\sqrt{\phi} \, j^{\mu},  \label{smu}\\
& & S^{a}=\sqrt{-\gamma}\sqrt{\phi} \, 
\Big(   -A_{\mu}^{\ a}j^{\mu} + j^{a} \Big),   \label{sa} \\
& &j^{\mu}=\gamma^{\mu\nu}
        \phi^{a b}D_{\nu}\phi_{a b},           \label{jeimu}\\
& &j^{a}=\phi^{a b} \gamma^{\mu\nu}
        \partial_{b}\gamma_{\mu\nu}.           \label{jeiei}
\end{eqnarray}
\noindent

One can easily recognize that this action 
is in a form of a (1+1)-dimensional field theory action.
In geometrical terms the above action can be understood 
as follows. 
$\gamma^{\mu\nu}\hat{ {\rm R}}_{\mu\nu}$
can be interpreted as the ``gauged'' scalar curvature of $M_{1+1}$,
since the diff$N_{2}$-valued gauge fields are coupled to
$\gamma_{\mu\nu}$ and $\hat{ \Gamma}_{\mu\nu}^{\ \ \alpha}$ 
in the formulae (\ref{gric}) and (\ref{ok}).
$\phi^{a c}{\rm R}_{a c}$ is the scalar curvature of $N_{2}$,
which is proportional to the Euler 
characteristics $\chi$ when integrated over $N_{2}$. 

The remaining terms in (\ref{big}) are the {\it extrinsic} terms, 
telling us how $M_{1+1}$ and $N_{2}$ are embedded into the
enveloping 4-dimensional spacetime. 
Each term in (\ref{big}) is 
manifestly diff$N_{2}$-invariant, and the $y^{a}$-dependence 
of each term is completely ``hidden'' in the Lie derivatives.
In this sense we may view the fibre space $N_{2}$ as 
a kind of ``internal'' space as in Yang-Mills theory.
Thus, the Einstein-Hilbert 
action is describable as a (1+1)-dimensional
Yang-Mills type gauge theory interacting with
scalar fields and (1+1)-dimensional non-linear sigma fields 
of generic types, with couplings to curvatures of two 2-surfaces. 
The associated Yang-Mills gauge symmetry is the diff$N_{2}$
symmetry. 

\end{section}

\begin{section}{Discussions}
\label{s5}

In this paper, we presented the KK formalism of general
relativity of generic 4-dimensional spacetimes, 
viewing the spacetime 
as a local product of the (1+1)-dimensional base manifold and 
the 2-dimensional fibre space. 
Within this framework, we made a decomposition
of a given 4-dimensional spacetime metric into sets of fields
which transform as a tensor field, gauge fields, and
scalar fields under the group of the diffeomorphisms on $N_{2}$.
 
In connection with issues of quantum gravity, this KK approach
has the following aspects which deserve further remarks.
For instance, solving the Einstein's constraint 
equations or constructing the gauge invariant physical observables 
is known to be one of the most important problems 
in quantum general relativity. 
In our formalism, the diffeomorphisms of 
the 2-dimensional space $N_{2}$ plays the role of 
a local gauge symmetry {\it exactly} as in Yang-Mills theory. 
Therefore the two constraint
equations associated with the diff$N_{2}$ transformations
can be ``automatically'' solved, 
using the diff$N_{2}$-invariant scalars. 
However, there are two additional 
constraint equations which require further studies
in order to fully take care of the four Einstein's 
constraint equations\cite{recent}.

It should be also stressed that the Lie derivative appears 
naturally
in this formalism, via the {\it minimal} couplings to the 
diff$N_{2}$-valued gauge fields. In the standard 
(3+1)-formalism, the natural derivative operator is the 
metric-compatible covariant derivative, which requires the metric 
be non-degenerate. The Lie derivative, on the other hand, can be
defined even when the metric is degenerate. For instance,
at null infinity ${\cal I}^{+}$ of the asymptotically 
flat spacetimes, the natural 
derivative operator is the Lie derivative, rather than the 
covariant derivative, because the metric 
on ${\cal I}^{+}$ is degenerate with the signature 
$(0,+,+)$\cite{geroch77}. Therefore, 
the KK formalism, based on the notion of the Lie derivative,
should be extendable to spacetimes where the metric is degenerate, 
which would be difficult to describe in conventional approaches.

Finally, it will be a challenging
problem to try to {\it reinterpret} the exact solutions
of the Einstein's equations from this gauge theory point of view.
This seems very interesting, for there are a number of exact
solutions of the Einstein's equations which do not permit sensible
physical interpretations from the 4-dimensional spacetime 
perspective\cite{exact80}.

\end{section}

\vskip 1.5cm
\noindent
\centerline{\bf Acknowledgments}\\

The author thanks the referee for informing him the related work 
of L.J. Mason and E.T. Newman [1]. 
This work is supported in part by Korea Science and Engineering 
Foundation (95-0702-04-01-3). 

\nopagebreak

\end{document}